\documentclass[10pt,twocolumn,letterpaper]{article}

\usepackage[utf8]{inputenc}
\usepackage[T1]{fontenc}
\usepackage[margin=0.75in,top=0.75in,bottom=1in]{geometry}
\usepackage{graphicx}
\usepackage{hyperref}
\usepackage{url}
\usepackage[english]{babel}
\usepackage{adjustbox}
\usepackage{tabularx}
\usepackage{subcaption}
\usepackage{amsmath}
\usepackage{booktabs}
\usepackage{titlesec}
\usepackage{abstract}

\titleformat{\section}{\normalfont\bfseries\scshape}{\Roman{section}.}{0.5em}{}
\titleformat{\subsection}{\normalfont\itshape}{\Alph{subsection}.}{0.5em}{}
\titlespacing*{\section}{0pt}{6pt}{3pt}
\titlespacing*{\subsection}{0pt}{4pt}{2pt}

\usepackage{tabularx}

\title{\LARGE \textbf{A Manual Bar-by-Bar Tempo Measurement Protocol for Polyphonic Chamber Music Recordings: Design, Validation, and Application to Beethoven's Piano and Cello Sonatas}}

\author{Dr Ignasi Sole \quad \texttt{ignasiphd@gmail.com} \quad \today}

\makeatletter
\renewcommand{\@biblabel}[1]{[#1]}
\renewcommand{\thebibliography}[1]{%
  \section*{\textbf{References}}
  \small
  \list{\@biblabel{\@arabic\c@enumiv}}{%
    \settowidth\labelwidth{\@biblabel{#1}}%
    \leftmargin\labelwidth
    \advance\leftmargin\labelsep
    \usecounter{enumiv}%
    \let\p@enumiv\@empty
    \renewcommand\theenumiv{\@arabic\c@enumiv}}%
  \sloppy\clubpenalty4000\widowpenalty4000%
  \sfcode`\.\@m}
\makeatother

\begin{document}

\maketitle
\thispagestyle{empty}
\pagestyle{empty}

%%%%%%%%%%%%%%%%%%%%%%%%%%%%%%%%%%%%%%%%%%%%%%%%%%%%%%%%%%%%%%%%%%%%%%%%%%%%%%%%

\begin{abstract}
Empirical performance analysis depends on the accurate extraction of tempo data from recordings, yet standard computational tools, designed for monophonic audio or modern studio conditions, fail systematically when applied to historical polyphonic chamber music. This paper documents the failure of automated beat-detection software on duo recordings of Beethoven's five piano and cello sonatas (Op.~5 Nos.~1 and~2; Op.~69; Op.~102 Nos.~1 and~2), and presents a formalised manual alternative: a cumulative lap-timer protocol that yields bar-level beats-per-minute data with millisecond resolution. The protocol, developed in cross-disciplinary collaboration with an engineer specialising in VLSI design, rests on a cumulative timestamp architecture that prevents error accumulation, permits internal self-validation, and captures expressive timing phenomena (rubato, fermatas, accelerandi, ritardandi) that automated tools systematically suppress or misread. The mathematical derivation of the BPM formula, the spreadsheet data structure, and the error characterisation are presented in full. Applied to over one hundred movement-level recordings spanning 1930--2012, the protocol generated a dataset subsequently visualised through tempographs, histograms with spline-smoothed probability density functions, ridgeline plots, and combination charts. The paper argues that manual annotation is not a methodological retreat but a principled response to the intrinsic limitations of computational tools when faced with the specific challenges of polyphonic historical recordings. The complete dataset and analysis code are publicly available.
\end{abstract}

%%%%%%%%%%%%%%%%%%%%%%%%%%%%%%%%%%%%%%%%%%%%%%%%%%%%%%%%%%%%%%%%%%%%%%%%%%%%%%%%

\section{Introduction}

The empirical analysis of recorded musical performance has matured substantially since Bowen's foundational work in 1996, which demonstrated that tempo, duration, and expressive flexibility leave measurable traces amenable to systematic study~\cite{c1}. Since then, the field has been shaped by increasingly powerful computational tools: Sonic Visualizer~\cite{c2}, the analytical platforms developed through CHARM~\cite{c3}, and a growing body of Music Information Retrieval (MIR) software capable of extracting onset times, beat grids, and tempo curves from audio with minimal human intervention. This infrastructure has enabled large-scale corpus studies of solo piano~\cite{c4}, orchestral~\cite{c1}, and vocal~\cite{c3} repertoire.

Yet a significant class of repertoire remains resistant to these tools: historical polyphonic chamber music recordings. Duo textures, in which two instruments of similar spectral range (such as cello and piano) sound simultaneously, pose a documented challenge for beat-tracking and onset-detection algorithms. The timbral overlap between instruments, combined with the degraded audio quality characteristic of recordings made before the 1950s, produces conditions under which automated analysis either fails outright or yields data of insufficient reliability for scholarly use.

This paper addresses that gap. It reports the systematic failure of automated analysis on a corpus of recordings of Beethoven's five piano and cello sonatas, presents a formalised manual alternative, and demonstrates its application and outputs. The contribution is methodological: not a finding about Beethoven performance practice in itself, but a validated protocol that other researchers may adopt and adapt when confronted with equivalent challenges.

The remainder of the paper is structured as follows. Section~II reviews the existing landscape of computational and manual tempo analysis methods, identifying the specific conditions under which each approach breaks down. Section~III documents the failure of automated tools on the target corpus. Section~IV presents the manual protocol in full, including its mathematical basis, data architecture, and error model. Section~V describes the spreadsheet data structure and BPM computation. Section~VI presents the visualisation suite generated from the resulting dataset. Section~VII discusses the protocol's limitations, reproducibility, and potential for extension. Section~VIII concludes.

%%%%%%%%%%%%%%%%%%%%%%%%%%%%%%%%%%%%%%%%%%%%%%%%%%%%%%%%%%%%%%%%%%%%%%%%%%%%%%%%

\section{Background: Existing Approaches to Tempo Extraction}

\subsection{Computational Methods and Their Assumptions}

Automated tempo extraction tools rest on a set of shared assumptions that are rarely stated explicitly but are consequential for their applicability. Beat tracking algorithms, the most widely used class of tool, assume a detectable periodic pulse in the audio signal. They perform well when the metric structure is regular, the recording is acoustically clear, and the primary melodic line is spectrally separable from any accompaniment. These conditions hold reasonably well for modern solo piano recordings, for orchestral music where the bass frequencies carry a clear metrical signal, and for popular music genres where the drum track is metrically dominant.

They do not hold for historical duo recordings. When a cello and piano sound together, the fundamental frequencies of the two instruments overlap substantially across much of the cello's range. The piano's attack transients, which beat-trackers rely on heavily, are masked by the cello's sustained tone. In recordings from the 1930s and 1940s, additional noise from lacquer disc surfaces, analogue-to-digital conversion artefacts, and microphone frequency response characteristics further degrade the signal. The result is that beat-trackers, when applied to such recordings, either produce erratic tempo estimates or fail to return meaningful data at all.

Dannenberg and Mohan, in their study of tempo characterisation across jazz and rock recordings, acknowledge that tapping-based methods produce unavoidable ``tap-timing jitter''~\cite{c5}. Their compensatory approach (multiple trials on a microphone in synchrony with playback) is workable for metrically regular genres but does not scale to the expressive freedom of classical chamber music, where performers routinely depart from a strict metronomic pulse through rubato, ritardandi, and fermatas. Cook similarly notes that tapping ``is not sufficiently accurate'' at the level of individual beats~\cite{c3}, pointing researchers toward simpler macro-timing approaches such as the stopwatch for large-scale duration measurements.

Repp's influential study of timing patterns in Beethoven piano music used automated onset detection on solo piano recordings but acknowledged substantial manual correction of the resulting data~\cite{c6}. This semi-automated approach is viable for solo piano but does not transfer to polyphonic duo textures where the two instrument lines cannot be cleanly separated.

\subsection{Manual and Semi-Manual Approaches in the Literature}

Manual timing approaches have a long history in performance research, predating computational tools entirely. McEwen's 1928 analysis of rubato in piano recordings used the physical lengths of perforations in Duo-Art piano rolls to measure note durations~\cite{c7}. Hartmann's 1932 study employed a Bösendorfer computer-monitored grand piano to measure asynchrony in pianists' performances of Chopin~\cite{c8}. Bowen's use of a stopwatch for large-scale duration timing represents a more recent continuation of this tradition~\cite{c1}, endorsed by Cook as appropriate for ``performance time'' measurements where individual bar resolution is not required~\cite{c3}.

Pati, Lerch, Arthur, and Gururani, in their interdisciplinary review of music performance analysis, observe that comparing multiple recordings of a single work across different performers and career stages is the most reliable method for approaching ``the nature of the performance parameter''~\cite{c9}. Their framework categorises performance elements into tempo, dynamics, pitch, and timbre, extracted through tools such as Sonic Visualizer and Audacity. Crucially, they note that pre-approval of recordings by performers implies intentionality, a claim that is clearly inapplicable to early twentieth-century recordings where musicians often had minimal control over production decisions~\cite{c9}.

What has been absent from the literature is a formalised manual protocol that operates at bar-level resolution, captures expressive timing in its full complexity, quantifies its own error characteristics, and is designed from the outset to be reproducible by other researchers. This paper presents such a protocol.

%%%%%%%%%%%%%%%%%%%%%%%%%%%%%%%%%%%%%%%%%%%%%%%%%%%%%%%%%%%%%%%%%%%%%%%%%%%%%%%%

\section{Failure of Automated Analysis on the Target Corpus}

The corpus under study consists of over one hundred recordings of individual movements from Beethoven's five piano and cello sonatas (Op.~5 No.~1, Op.~5 No.~2, Op.~69, Op.~102 No.~1, Op.~102 No.~2), spanning the period 1930--2012. Recordings were sourced from commercial releases and digitised archival collections. The earliest recordings in the corpus were made under conditions of severe technical constraint: limited frequency response, surface noise, and the necessity of continuous performance without editing.

Automated beat extraction was attempted using MUsanim (the Music Animation Machine toolkit), chosen for its use in several musicological corpus studies. The goal was to detect note onsets or beat positions from the polyphonic audio and derive a bar-level tempo curve that could be compared across recordings. This approach failed to yield meaningful data. Across the corpus, the software produced one of three failure modes: complete inability to detect a beat grid; a grid locked to the piano's attack transients that ignored the cello's contribution to the metric structure; or erratic tempo estimates fluctuating by hundreds of BPM within single bars, reflecting the algorithm's misidentification of overtones and noise artefacts as onset events.

These failures are not idiosyncratic to MUsanim. They reflect structural limitations of onset-detection algorithms when applied to polyphonic audio with a high degree of spectral overlap between instruments. The piano's sustain pedal, used extensively in Beethoven's chamber music, further blurs attack transients that algorithms depend on to place beat positions. Early recordings, particularly those made before 1950, compound these difficulties through analogue noise and frequency-range compression.

The conclusion was unambiguous: automated tempo extraction was not a viable primary method for this corpus. A manual alternative was required.

%%%%%%%%%%%%%%%%%%%%%%%%%%%%%%%%%%%%%%%%%%%%%%%%%%%%%%%%%%%%%%%%%%%%%%%%%%%%%%%%

\section{The Manual Bar-by-Bar Protocol}

\subsection{Design Principles and Cross-Disciplinary Development}

The manual protocol was developed in collaboration with Jordi Altay\'o, a doctoral researcher at KTH Royal Institute of Technology specialising in VLSI design. Although Altay\'o's expertise lies in electronic engineering rather than musicology, this cross-disciplinary context proved methodologically productive: it brought an emphasis on accuracy, error modelling, and data organisation that might not have emerged from within a purely musicological framework. The collaboration exemplifies a broader principle; that performance analysis benefits from engagement with fields where precision data collection and measurement error are treated as first-class methodological concerns.

The protocol was designed around four principles. First, \textit{cumulative measurement}: timestamps are recorded as cumulative elapsed time from the movement's start, not as individual bar durations. Second, \textit{error isolation}: timing errors at any single bar affect only that bar's duration, not subsequent bars. Third, \textit{internal self-validation}: the sum of all bar durations must equal the independently measured total movement duration. Fourth, \textit{expressive fidelity}: the method tracks what performers actually did, including fermatas, rubato, and structural tempo shifts, rather than what the score prescribes.

\subsection{Data Collection Procedure}

The primary measurement instrument was a digital stopwatch with lap-timer functionality. The Stopwatch: StopNow Free application was used on Windows; any equivalent application that exports lap data in CSV format may be substituted on other platforms, including macOS and iOS.

For each recording and each movement, the following procedure was observed. The annotator began listening to the audio recording with the score open. The stopwatch was started precisely at the moment the first bar begins; the downbeat of the movement, not the beginning of any anacrusis, which was timed separately. As the recording progressed, the lap button was pressed at each barline: the moment the notated measure count advanced by one. The stopwatch was not paused or reset between bars; it ran continuously throughout the movement. Each lap press therefore recorded the cumulative elapsed time from the movement's start to the end of that bar. The final lap recorded the moment the movement concluded.

Prior to the main data collection, the annotator practised with multiple recordings to calibrate reaction time and develop familiarity with each movement's score. This rehearsal phase reduced systematic timing bias and ensured that the annotator could follow the score reliably even through passages of significant tempo flexibility.

\subsection{Fermata and Ambiguous Barline Handling}

Fermatas and extended pauses posed a specific challenge. When a fermata or negotiated pause between performers obscured the transition between bars, the following procedure was applied. The passage was first listened to without annotation to identify the onset of the bar following the fermata. The annotator then replayed the passage and timed the barline onset explicitly. The measurement was verified by a second independent replay. If the two measurements diverged by more than 0.2 seconds, a third measurement was taken and the two closest values averaged.

This procedure treats the fermata as a boundary event between two timed segments rather than as a duration to be measured in itself. It reflects the analytical reality that fermata lengths in chamber music are negotiated between performers in real time and are not determinable from the score alone. The resulting data therefore captures the performed temporal structure faithfully, including the expressive weight that performers assign to structural pauses.

\subsection{Mathematical Basis of the Protocol}

Let a movement consist of $M$ bars. Applying the lap procedure yields a sequence of cumulative timestamps

\[
T_1,\; T_2,\; \ldots,\; T_M
\]

where $T_i$ denotes the total elapsed time from the movement's start to the end of bar $i$, and $T_M$ is therefore the total movement duration. Setting $T_0 = 0$ (the movement's downbeat), the duration of bar $i$ is

\begin{equation}
\Delta t_i = T_i - T_{i-1}, \quad i = 1, 2, \ldots, M.
\label{eq:barduration}
\end{equation}

Letting $n_i$ denote the number of beats in bar $i$ (derived from the time signature, adjusted for any notated meter change or anacrusis), the instantaneous tempo for bar $i$ in beats per minute is

\begin{equation}
\mathrm{BPM}_i = \frac{n_i \times 60}{\Delta t_i},
\label{eq:bpm}
\end{equation}

where $\Delta t_i$ is expressed in seconds. Because timestamps are stored in spreadsheet software as decimal days (the internal representation used by both Microsoft Excel and Google Sheets), the conversion factor of 86,400 seconds per day must be applied:

\begin{equation}
\mathrm{BPM}_i = \frac{n_i \times 60}{\Delta t_i^{(\mathrm{days})} \times 86400}.
\label{eq:bpm_days}
\end{equation}

\textit{Internal consistency check.} The self-validating property of the cumulative architecture follows directly from the definition:

\begin{equation}
\sum_{i=1}^{M} \Delta t_i = T_M.
\label{eq:consistency}
\end{equation}

In spreadsheet implementation, this means that the sum of all bar duration values in the $\Delta t$ column must equal the total movement duration recorded at $T_M$. Any discrepancy exceeding the expected reaction-time tolerance indicates a missed lap press or a data-entry error. This check was applied to every movement in the corpus before the data was accepted for further analysis.

\textit{Example.} For a bar in common time ($n = 4$) with a measured cumulative timestamp difference of $\Delta t = 1.708 \times 10^{-5}$ days (approximately 1.476 seconds):

\[
\mathrm{BPM} = \frac{4 \times 60}{1.476} \approx 162.7 \;\text{BPM}.
\]

\subsection{Error Characterisation}

Human reaction time introduces a timing uncertainty of approximately $\pm 0.1$ seconds per lap press. The impact of this error on the BPM calculation for a single bar can be estimated by differentiating Equation~(\ref{eq:bpm}) with respect to $\Delta t$:

\[
\frac{\partial \;\mathrm{BPM}}{\partial \;\Delta t} = -\frac{n \times 60}{\Delta t^2}.
\]

For a bar lasting 1.5 seconds at approximately 160 BPM in $4/4$ time, a timing error of $\delta t = 0.1$ seconds produces a BPM error of

\[
|\delta\;\mathrm{BPM}| = \frac{4 \times 60}{(1.5)^2} \times 0.1 \approx 10.7 \;\text{BPM}.
\]

Three properties of this error are critical for assessing its methodological significance. First, the error is \textit{random rather than systematic}: reaction-time delays have no directional bias, so they do not shift the tempo estimate consistently upward or downward. Second, the error is \textit{non-accumulating}: because each bar's duration is computed as the difference between two consecutive cumulative timestamps, a delayed press at bar $i$ affects only $\Delta t_i$ and $\Delta t_{i+1}$ (the latter being shortened by the same amount the former was extended), and propagates no further. Third, the error is \textit{smoothed by aggregation}: for section-level or movement-level averages over many bars, random errors cancel and the aggregate BPM converges toward the true value. The impact of single-bar errors on the comparative analyses presented in subsequent chapters is therefore small and does not affect the direction or significance of observed trends.

The reaction-time error is also substantially smaller than the tempo variability observed between performers, which regularly exceeds 20--40 BPM across the corpus. The measurement noise is thus well below the signal being studied.

%%%%%%%%%%%%%%%%%%%%%%%%%%%%%%%%%%%%%%%%%%%%%%%%%%%%%%%%%%%%%%%%%%%%%%%%%%%%%%%%

\section{Data Structuring and BPM Computation}

\subsection{Spreadsheet Architecture}

Following data collection, cumulative lap times were exported from the stopwatch application as a CSV file and imported into a spreadsheet workbook (Google Sheets was used throughout; Microsoft Excel is fully equivalent). Each musical movement was allocated to a separate worksheet, reflecting the different time signatures and metric structures of individual movements. Within each worksheet, a consistent column structure was applied:

\begin{itemize}
    \item \textbf{Column A}: Bar number ($i = 1, 2, \ldots, M$), serving as the primary alignment key across recordings.
    \item \textbf{Column B}: Cumulative timestamp $T_i$ in decimal days.
    \item \textbf{Column C}: Bar duration $\Delta t_i$, computed as \texttt{=B\textit{n}-B\textit{(n-1)}}.
    \item \textbf{Column D}: Beats per bar $n_i$, derived from the time signature (adjusted at meter changes and anacrusis bars).
    \item \textbf{Column E}: BPM$_i$, computed from Equation~(\ref{eq:bpm_days}).
\end{itemize}

This five-column unit was repeated side-by-side for each recording, with bar numbers in Column A serving as the common index. The alignment by bar number (rather than by elapsed time or note count) is the methodological backbone of comparative performance analysis: it ensures that tempo values from different recordings are contextually matched to the same musical moment in the score.

\subsection{Multi-Recording Layout and Quality Control}

The master workbook contained one worksheet per movement across all five sonatas. Within each worksheet, the five-column block was replicated for every recording, with performer names and recording dates in a header row. This layout allowed immediate visual inspection of alignment: if one recording contained an extra bar due to a written-out repeat or a cut, the misalignment was immediately visible as a row-offset, prompting investigation and correction before analysis.

Quality control was applied to every movement. Two checks were performed: the consistency check from Equation~(\ref{eq:consistency}), verifying that $\sum \Delta t_i = T_M$; and a musical plausibility check, verifying that the BPM values in the tempo column corresponded to the annotator's listening experience of the recording. Where an anomalous BPM spike appeared (suggesting a missed or double-pressed lap) the recording was revisited and the lap timestamp corrected. The transparency of the spreadsheet layout made this correction straightforward: each anomaly was visible in context, flanked by the BPM values of the preceding and following bars.

\subsection{Dataset Availability}

The complete tempo dataset, including cumulative timestamps, bar durations, and computed BPM values for all analysed recordings, is publicly available at:

\begin{center}
\url{https://github.com/isolepinas/PhD-Appendix/tree/main/Tempo\%20Dataset}
\end{center}

%%%%%%%%%%%%%%%%%%%%%%%%%%%%%%%%%%%%%%%%%%%%%%%%%%%%%%%%%%%%%%%%%%%%%%%%%%%%%%%%

\section{Visualisation Suite}

The bar-level BPM dataset generated by the protocol was designed from the outset to support a suite of complementary visualisations, each revealing a different dimension of interpretive behaviour. Five visualisation types were developed and applied systematically across the corpus. This section describes each in turn.

\subsection{Tempographs}

A tempograph is a line chart plotting BPM against bar index, producing a continuous tempo curve for each recording. Tempographs are the primary tool for close reading of interpretive decisions: they capture rubato, accelerandi, ritardandi, and structural tempo contrasts at bar-level resolution. Overlaying multiple tempographs in a single figure enables direct comparison of interpretive strategies at specific structural moments.

\begin{figure}[htpb]
  \centering
  \includegraphics[width=\columnwidth]{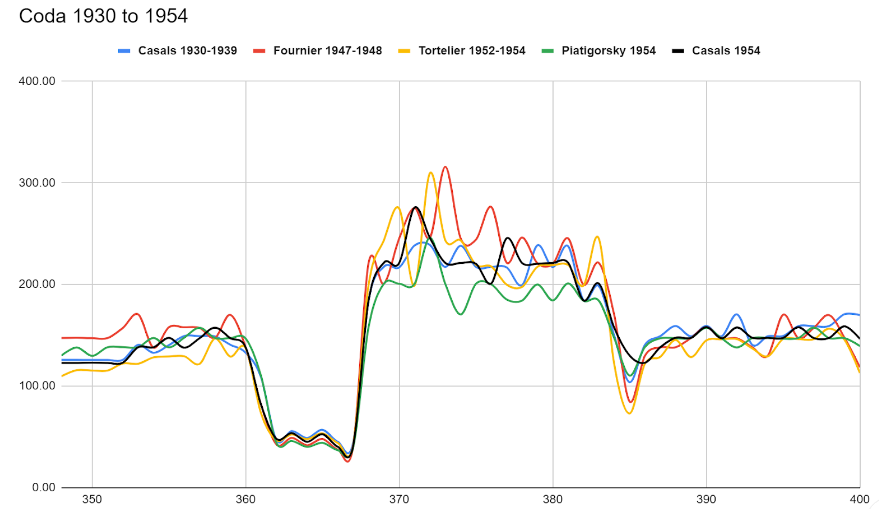}
  \caption[Tempo graph of mm.\,350–400 from Sonata Op.\,5 No.\,1 by L.\,V.\,Beethoven.]{Overlay of five representative cellists' tempo curves (BPM vs. bar index) for mm.\,350–400 of Sonata Op.\,5 No.\,1, First Movement, with phrase boundaries at mm.\,355 and 380 indicated.}
  \label{fig:tempograph350-400}
\end{figure}

To avoid visual crowding when displaying large numbers of recordings, two complementary formats were employed. Focused tempographs overlaid no more than five selected recordings across spans of up to one hundred bars, enabling detailed comparative analysis of specific passages. Grid-format small-multiple arrays displayed one tempograph per recording across the full movement, sacrificing detail for comprehensiveness and allowing rapid scanning across the entire corpus.

\subsection{Histograms with Spline-Smoothed PDFs}

Histograms of the BPM distribution for each recording were generated in MATLAB. Each histogram shows the relative frequency of tempo values across all bars of a movement, providing a compact summary of interpretive pacing: a narrow histogram with a sharp peak indicates metronomic consistency; a broad or multimodal histogram indicates tempo flexibility or structural contrast.

To facilitate comparison of distributional shape across recordings, a smooth probability density function was superimposed on each histogram. The smoothing procedure operated as follows. The empirical cumulative distribution function (CDF) was computed by cumulative summation and normalisation of the histogram bin counts. A cubic spline was then fitted to this empirical CDF, with zero-slope boundary conditions enforced at the data's minimum and maximum tempo values to prevent artificial extrapolation. Differentiating the fitted spline yielded a continuous, smooth PDF. This spline-derived PDF was plotted in red over the histogram, enabling identification of the modal tempo and subtle distributional features (slight skewness, secondary humps) that raw histogram bins may obscure.

\begin{figure}[htp!]
  \centering
  \includegraphics[width=\columnwidth]{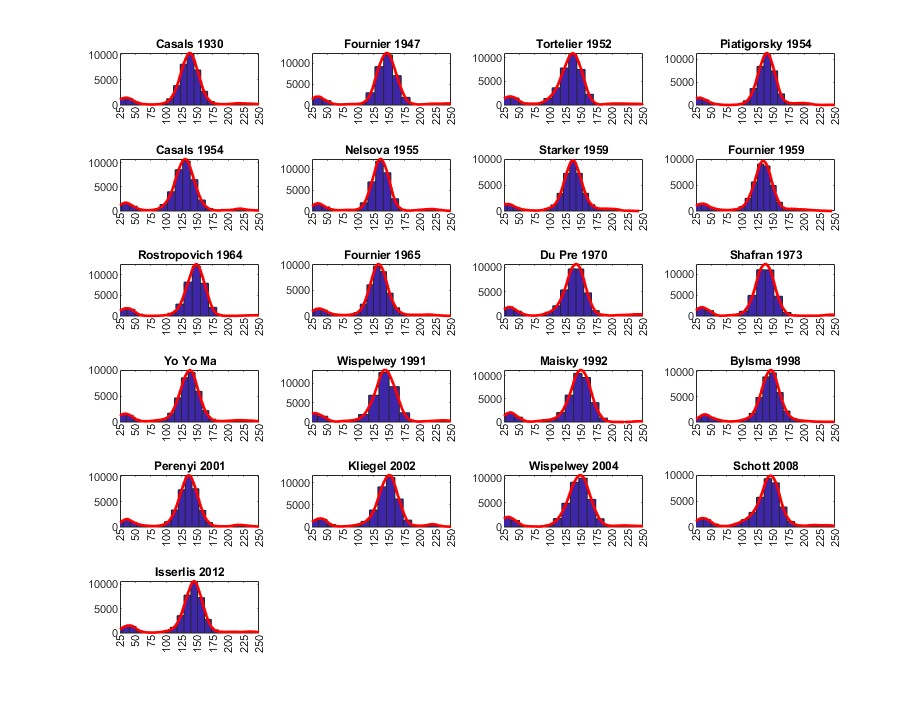}
  \caption[Histogram of tempo flexibility in Beethoven Sonata Op. 5 No. 1]{Per-recording histograms of beat-by-beat BPM for Sonata Op. 5 No. 1, First Movement. The X-axis shows tempo in 2 BPM bins; the Y-axis shows relative frequency of occurrence.}
  \label{fig:histogram_sonata1}
\end{figure}
\subsection{Ridgeline Plots of Tempo Distributions}

To compare tempo distributions across the full set of recordings within a single movement simultaneously, ridgeline plots (also known as joy plots) were employed~\cite{c10}. A ridgeline plot stacks kernel density estimates (KDE) vertically, one per recording, offset along the y-axis to prevent overlap. Each ridge represents the tempo distribution of a single performance. The horizontal axis is shared across all ridges, so the same tempo value aligns vertically through the figure, enabling direct visual comparison of distributional peaks, widths, and skewness.

Ridgeline plots were generated in Python using the Seaborn and Matplotlib libraries~\cite{c11, c12}. Data were loaded from CSV exports of the master spreadsheet using Pandas~\cite{c13}. Colour coding was applied to each ridge according to its recording's mean tempo, using a coolwarm palette: cooler colours (blue-green) indicated slower average tempi; warmer colours (orange-red) indicated faster average tempi. This colour encoding adds an additional analytical layer, allowing historical tempo trends to be read directly from the visual structure of the figure.

\begin{figure}[htp!]
  \centering
  \includegraphics[width=\columnwidth]{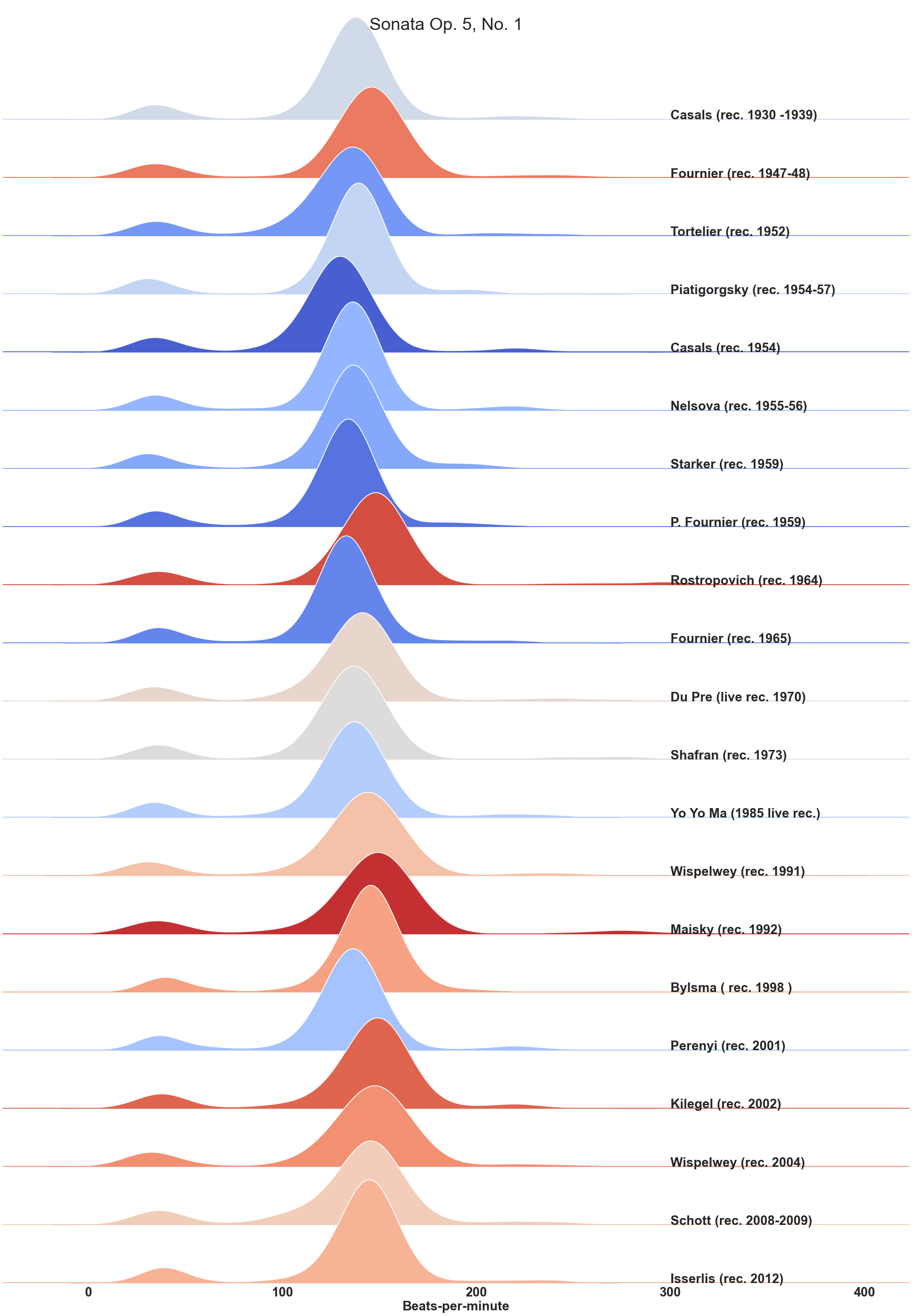}
  \caption{Ridge plot of per-performance tempo distributions (KDE) for Op.\,5 No.\,1, First Movement, annotated by performer and year.}
  \label{fig:tempoRidge}
\end{figure}

\subsection{Stacked Bar Charts and Combination Charts}

Stacked bar charts were used to illustrate how each performer distributed overall performance duration across the formal sections of a movement. Each bar represents a single recording; each coloured segment within the bar represents one formal section's duration as a proportion of the total movement duration. This layout enables immediate identification of performers who assign disproportionate temporal weight to specific sections, for example, a notably extended development section, relative to the corpus norm.

Combination charts overlaid average BPM per recording (displayed as filled bars against a primary y-axis) with tempo variability (standard deviation of bar-level BPM, displayed as a line graph against a secondary y-axis). Historical metronome reference marks from Czerny, Moscheles, and Kolisch were superimposed as horizontal reference lines. This multi-layer format allowed simultaneous comparison of absolute tempo, internal flexibility, and deviation from historical recommendations across the full corpus.

Table~\ref{table:viz_summary} summarises the analytical properties of the five visualisation types. A perfectly metronomic performance would manifest as a flat tempograph, a narrow histogram, a thin ridgeline ridge, and minimal variability in combination charts. Performances characterised by expressive rubato manifest as the opposite across all five formats. The consistency of these signals across multiple visualisation types provides cross-validation of interpretive conclusions.

\begin{table}[htbp]
\caption{Visualisation Types and Their Analytical Properties}
\centering
\begin{tabularx}{\columnwidth}{|X|X|X|}
\hline
\textbf{Visualisation} & \textbf{Granularity} & \textbf{Primary Use} \\
\hline
Tempograph & Bar-level & Close reading of tempo profile \\
\hline
Histogram + spline PDF & Distribution & Modal tempo and spread \\
\hline
Ridgeline plot & Distribution (multi-recording) & Cross-corpus comparison \\
\hline
Stacked bar chart & Section-level & Structural pacing \\
\hline
Combination chart & Recording-level & Tempo, variability, and historical benchmarks \\
\hline
\end{tabularx}
\label{table:viz_summary}
\end{table}

%%%%%%%%%%%%%%%%%%%%%%%%%%%%%%%%%%%%%%%%%%%%%%%%%%%%%%%%%%%%%%%%%%%%%%%%%%%%%%%%

\section{Discussion}

\subsection{The Protocol as a Principled Response to Computational Limits}

The decision to adopt a manual approach was not a retreat from methodological rigour but its opposite. The automated tools that failed on this corpus were designed for conditions that historical polyphonic chamber recordings systematically violate. Using them regardless, and accepting their erratic output, would have produced data of lower reliability than the manual alternative, while concealing the unreliability behind the appearance of algorithmic authority. The manual protocol, by contrast, makes its error model explicit, quantifies the magnitude and distribution of its errors, and provides a self-validation mechanism. It is, in this sense, more epistemically transparent than the computational tools it replaces.

This is consistent with a broader principle articulated across the performance analysis literature: computational tools should complement, not replace, the human and historical dimensions of musical interpretation~\cite{c9}. Katz's observation that performers adjust their interpretations in real time during recording sessions~\cite{c14}, an observation clearly inapplicable to the assumption of a fixed, recoverable ground truth built into automated tools, underscores the importance of analytical methods that remain sensitive to the contingent, intentional character of recorded performance.

\subsection{The Human-in-the-Loop Advantage}

The manual protocol preserves information that automated systems discard. A musically trained annotator following the score can identify and correctly timestamp structural events; the arrival of a new theme, the onset of a development section, the moment a fermata resolves, that have no acoustic signature detectable to a beat-tracking algorithm. The annotator's musical knowledge is not a source of bias but a prerequisite for the data's musical meaningfulness.

This advantage is particularly significant for the corpus under study. Beethoven's piano and cello sonatas contain passages of extreme tempo flexibility: the slow introductions to Op.~5 Nos.~1 and~2, the rhapsodic character of Op.~102 No.~1's opening, the long-sustained adagio passages in Op.~69. These passages are precisely the moments where automated tools fail most completely and where the interpretive decisions under study are most consequential. The manual protocol captures them; automated tools cannot.

\subsection{Scalability and Limitations}

The most significant limitation of the protocol is the time it demands. Timing a single movement at bar-level resolution with the care required by the quality control procedure described in Section~IV takes approximately 30--45 minutes per recording for a movement of average length. Applied to over one hundred movement-level recordings, the data collection phase required several hundred hours of focused work. This investment is feasible for a doctoral study but limits the scalability of the approach to the very large corpora now common in computational performance analysis.

A hybrid approach may offer the most practical path forward. For repertoire and recording conditions where automated tools perform reliably (modern solo piano recordings, orchestral music with a strong bass pulse, recordings with high signal-to-noise ratios), automated extraction remains appropriate and efficient. The manual protocol presented here is reserved for cases where automated tools are known to fail: historical polyphonic recordings, degraded audio, expressive styles that systematically depart from metronomic regularity. Researchers undertaking new studies should assess their corpus against these criteria before choosing a data collection strategy.

A second limitation concerns inter-rater reliability. The current study was conducted by a single annotator. While the internal consistency check provides a self-validation mechanism, it does not address the question of how closely the protocol's outputs would be replicated by a different trained annotator. Future work should establish inter-rater reliability by having an independent musician time a representative sample of recordings and comparing the resulting BPM values bar by bar. The anticipated result (high agreement on structurally clear passages, moderate disagreement at fermatas and rapid tempo transitions) would be informative both about the protocol's reliability and about the genuine ambiguity of performed metre in these works.

\subsection{Reproducibility and Open Data}

All data, code, and supplementary materials generated by this study are publicly available. The complete BPM dataset is archived at the GitHub repository cited in Section~V. The Python code for ridgeline plots and the MATLAB code for histograms are available in the same repository under \texttt{Coded Graphs/}. The annotated portamento scores are available under \texttt{Portamento Scores/}. This openness reflects a commitment to reproducibility that is still relatively rare in musicological research but is increasingly standard in adjacent fields. It means that other researchers can extend the dataset with new recordings, challenge specific measurements, or apply the visualisation pipeline to different repertoire.

%%%%%%%%%%%%%%%%%%%%%%%%%%%%%%%%%%%%%%%%%%%%%%%%%%%%%%%%%%%%%%%%%%%%%%%%%%%%%%%%

\section{Conclusion}

This paper has presented a formalised manual protocol for bar-level tempo measurement of polyphonic chamber music recordings, developed in response to the systematic failure of automated beat-detection tools on historical duo recordings. The protocol's key contributions are: a cumulative timestamp architecture that prevents error accumulation and enables internal self-validation; a full mathematical derivation of the BPM formula; a quantified error model demonstrating that reaction-time uncertainty is random, non-accumulating, and small relative to the interpretive variation under study; and a five-type visualisation suite that transforms bar-level BPM data into analytical representations suited to different research questions.

Applied to over one hundred movement-level recordings of Beethoven's five piano and cello sonatas spanning 1930--2012, the protocol generated a dataset of sufficient resolution to trace bar-by-bar interpretive decisions, compare them across performers and generations, and situate them against historical metronome recommendations by Czerny, Moscheles, and Kolisch. The resulting dataset is publicly available and designed to be extended.

The broader argument of this paper is that methodological transparency (including the honest acknowledgement of a tool's failure and the principled design of an alternative) is a form of scholarly rigour. Computational tools are powerful precisely because they are fast and scalable; they are limited precisely because their assumptions are fixed. When the repertoire and the historical conditions fall outside those assumptions, the appropriate response is not to force the data through an inappropriate pipeline but to design a method suited to the problem. That is what the protocol presented here represents.

%%%%%%%%%%%%%%%%%%%%%%%%%%%%%%%%%%%%%%%%%%%%%%%%%%%%%%%%%%%%%%%%%%%%%%%%%%%%%%%%


\begin{thebibliography}{99}

    \bibitem{c1} J.~A.~Bowen, ``Tempo, Duration and Flexibility: Techniques in the Analysis of Performance,'' \textit{Journal of Musicological Research}, vol.~16, pp.~111--156, 1996.

    \bibitem{c2} C.~Cannam, C.~Landone, and M.~Sandler, ``Sonic Visualiser: An Open Source Application for Viewing, Analysing, and Annotating Music Audio Files,'' in \textit{Proceedings of the ACM International Conference on Multimedia}, Florence, 2010, pp.~1467--1468.

    \bibitem{c3} N.~Cook and D.~Leech-Wilkinson, ``A Musicologist's Guide to Sonic Visualizer,'' CHARM, 2009. [Online]. Available: \url{https://charm.rhul.ac.uk/analysing/p9_1.html}

    \bibitem{c4} N.~Cook, ``Performance Analysis and Chopin's Mazurkas,'' \textit{Musicae Scientiae}, vol.~11, no.~2, pp.~183--207, 2007.

    \bibitem{c5} R.~Dannenberg and S.~Mohan, ``Characterizing Tempo Change in Musical Performances,'' in \textit{Proceedings of the International Computer Music Conference 2011}, University of Huddersfield, July 31--August 5, 2011, pp.~650--656.

    \bibitem{c6} B.~Repp, ``Patterns of Expressive Timing in Performances of a Beethoven Minuet by Nineteen Famous Pianists,'' \textit{Journal of the Acoustical Society of America}, vol.~88, no.~2, pp.~622--641, 1990.

    \bibitem{c7} J.~B.~McEwen, \textit{Tempo Rubato or Time-Variation in Musical Performance}. Oxford: Oxford University Press, 1928.

    \bibitem{c8} W.~Goebl, \textit{Melody Lead in Piano Performance: Expressive Device or Artifact?} Vienna: Austrian Research Institute for Artificial Intelligence, 2001, citing A.~Hartmann, ``Untersuchungen \"{u}ber das metrische Verhalten in musikalischen Interpretation Varianten,'' \textit{Archiv f\"{u}r die gesamte Psychologie}, vol.~84, pp.~103--192, 1932.

    \bibitem{c9} A.~Pati, A.~Lerch, C.~Arthur, and S.~Gururani, ``An Interdisciplinary Review of Music Performance Analysis,'' \textit{Transactions of the International Society for Music Information Retrieval}, vol.~3, pp.~221--245, 2020. DOI: 10.5334/tismir.53.

    \bibitem{c10} B.~W.~Silverman, \textit{Density Estimation for Statistics and Data Analysis}. London: Chapman \& Hall, 1986.

    \bibitem{c11} J.~D.~Hunter, ``Matplotlib: A 2D Graphics Environment,'' \textit{Computing in Science \& Engineering}, vol.~9, no.~3, pp.~90--95, 2007.

    \bibitem{c12} M.~L.~Waskom, ``Seaborn: Statistical Data Visualization,'' \textit{Journal of Open Source Software}, vol.~6, no.~60, p.~3021, 2021.

    \bibitem{c13} W.~McKinney, ``Data Structures for Statistical Computing in Python,'' in \textit{Proceedings of the 9th Python in Science Conference}, S.~van der Walt and J.~Millman, Eds. Austin, TX: SciPy, 2010, pp.~51--56.

    \bibitem{c14} M.~Katz, \textit{Capturing Sound: How Technology Has Changed Music}. Berkeley: University of California Press, 2004.

    \bibitem{c15} M.~Noorduin, \textit{Beethoven's Tempo Indications}. PhD dissertation, University of Manchester, 2016. [Online]. Available: \url{https://www.escholar.manchester.ac.uk/uk-ac-man-scw:302884}

    \bibitem{c16} R.~Kolisch, ``Tempo and Character in Beethoven's Music,'' \textit{The Musical Quarterly}, vol.~77, no.~1, pp.~90--131, Spring 1993.

    \bibitem{c17} D.-J.~Povel, ``Temporal Structure of Performed Music: Some Preliminary Observations,'' \textit{Acta Psychologica}, vol.~41, pp.~309--320, 1977.

    \bibitem{c18} R.~Philip, \textit{Early Recordings and Musical Style: Changing Tastes in Instrumental Performance, 1900--1950}. Cambridge: Cambridge University Press, 1992.

    \bibitem{c19} D.~Leech-Wilkinson, \textit{The Changing Sound of Music: Approaches to Studying Recorded Musical Performance}. London: CHARM, 2009. [Online]. Available: \url{https://www.charm.kcl.ac.uk/studies/chapters/chap5.html}

    \bibitem{c20} T.~Kluyver et al., ``Jupyter Notebooks, a Publishing Format for Reproducible Computational Workflows,'' in \textit{Positioning and Power in Academic Publishing: Players, Agents and Agendas}, F.~Loizides and B.~Schmidt, Eds. Amsterdam: IOS Press, 2016, pp.~87--90.

\end{thebibliography}
\end{document}